\documentclass[lettersize,journal]{IEEEtran}
\usepackage{amsmath,amsfonts}
\usepackage{algorithmic}
\usepackage{algorithm}
\usepackage{array}
\usepackage[caption=false,font=normalsize,labelfont=sf,textfont=sf]{subfig}
\usepackage{textcomp}
\usepackage{stfloats}
\usepackage{url}
\usepackage{verbatim}
\usepackage{graphicx}
\usepackage{cite}

\usepackage{fancyhdr}

\usepackage{xcolor}
\usepackage{soul}

\usepackage{tikz}

\usepackage{multirow}
\usepackage{tabularx}
\usepackage{booktabs}

\usepackage{pifont}
\newcommand{\xmark}{\ding{55}}

\begin{document}

\title{Agent\textit{xG}Core: Agentic AI for Next-Generation Mobile Core Network}
% Towards AI-Native Next-Generation Core Networks for the Agentic AI Era
\author{Katarine Santana and Kelvin L. Dias\\
Centro de Informática - Universidade Federal de Pernambuco, Brazil\\
E-mail: \{mksb, kld\}@cin.ufpe.br

\thanks{Maria K. S. Barbosa and Kelvin L. Dias are with the Centro de Informática - Universidade Federal de Pernambuco, 50740-560, Recife-PE, Brasil (e-mail: mksb@cin.ufpe.br; kld@cin.ufpe.br; \textbf{This paper has been accepted for publication in IEEE Network.}.}% <-this % stops a space
}

\maketitle

\begin{abstract}
To meet the stringent requirements of emerging applications and the increasingly complex network management and operation, the Next Generation Mobile Networks (NextG), or 6G, will adopt an AI-native architecture on the Core Network (CN). In this movement, the Third Generation Partnership Project (3GPP) has extended the cellular CN with new function as a first step toward integrating analytics, Artificial Intelligence (AI), and machine learning. However, those new functionalities are constrained by a centralized approach and managerial complexity. Furthermore, with the rise of Large Language Models (LLMs), a new era in network orchestration and management begins, leveraging and empowering the Intent-based Networking (IBN) paradigm. In addition, AI agents and Agentic AI integrate Reasoning and Acting (ReAct), enabling the usage of such intents to continuously interact with the network. Unlike state-of-the-art approaches that primarily employ Agentic AI to mitigate deployment and configuration complexity in the CN, this paper introduces Agent\textit{xG}Core, which leverages an Agentic AI-Native layer to extend the 3GPP architecture and enable a system based on the existing APIs across the Beyond Next Generation Core (xGC) domain. This proposal establishes an AI-driven closed-loop for continuous optimization based on real-time information, enabling self-organization and self-adaptation. Our approach involves a multi-agent specialized system, divided into a network planner agent, capable of visualizing the network state and developing a plan to meet the intents, and a network executor, responsible for criticizing and executing the plan. To validate the proposed solution, an environment was built using an open-source CN, heterogeneous datasets, and different LLMs were employed to demonstrate its effectiveness.

\end{abstract}


\section{Introduction} 

The advent of sixth-generation (6G) mobile networks aims to enable applications for immersive experiences, extended reality, holographic telepresence, sensory monitoring, unmanned vehicles, and the Internet of Everything. These applications impose increasingly stringent requirements on mobile networks, such as ultra-high data rates, hyper-reliable and low-latency communications, ubiquitous connectivity, and massive access \cite{challenges_6g}. 

The core network (CN), which is responsible for essential mobile network functionalities such as mobility management, connectivity, session management, policy control, and data analytics, is expected to remain a critical element of the 6G architecture, evolving from the current 5G Core (5GC) cloud-native and service-based architecture (SBA). Consequently, the role of the 6G Core (6GC) will become increasingly critical due to the need to orchestrate, manage, and optimize possibly an even larger and more heterogeneous set of Network Functions (NFs).

Therefore, management and orchestration are becoming increasingly complex and difficult to perform manually or semi-automatically. In this context, towards 6G networks, the Next Generation Mobile Networks (NGMN) Alliance highlights the importance of prioritizing operational aspects, such as managing increasing complexity, improving efficiency, and reducing costs. To address these challenges, the adoption of an integrated hyperautomation framework becomes essential, enabling the automation of the lifecycle of services, networks, and business domains, increasing ecosystem resilience and reducing operational costs, with a strong reliance on fully integrated Artificial Intelligence (AI) functionalities \cite{ngmn_6g_vision}.

To meet such demands, 6G will adopt an AI-native architecture capable of real-time decision-making, self-organization, and continuous self-optimization. Consequently, new paradigms in mobile networks have emerged, such as AI for Network (AI4NET), which adopts AI to improve performance and is applied throughout the core, including models for anomaly detection, traffic prediction, and load balancing. However, implementing and managing these traditional models in the network largely depends on the network operator, and they are usually applied to smaller, isolated tasks, often leading to conflicts in decision-making.

% Generative AI and Intent-Based Network
Moreover, with the explosion of Generative AI (GenAI), especially after the release of ChatGPT at the end of 2022, new opportunities for automating and autonomously operating telecom networks have begun, particularly in the application of Large Language Models (LLMs) for both planning and management. This growing attention toward GenAIs also initiated a new era of network management and orchestration, empowering the application of the Intent-Based Networking (IBN) paradigm \cite{ibm_ibn}. 

IBN allows network operators to define desired intents, and the network will self-organize to fulfill these requirements, whether orchestrating the network to meet Quality of Service (QoS) requirements for different applications, ensuring Service Level Agreement (SLA), allocating resources for network slices, or even improving network energy efficiency. Based on the intent, the LLM can generate and execute action plans. In the state of the art, some studies have used GenAI in conjunction with the IBN concept for 5G core deployment \cite{ibn_5g_deploy}, service orchestration \cite{ibm_llm}, and NF management \cite{ibn_service}.

Nevertheless, these approaches still rely heavily on user intervention and fail to address QoS-driven intents for the 5G and beyond core. As a result, current solutions are insufficient to cope with the demands of emerging traffic categories and the growing number of NFs, which require continuous optimization and autonomous multi-step decision-making, planning, and coordination within the Beyond Next Generation Core (xGC) domain, involving the Management and Network Orchestration (MANO) framework and the CN control and data planes.

Moreover, GenAI models alone are inherently limited, as they cannot directly interact with the operational network. This limitation has motivated the integration of Reasoning and Acting (ReAct), enabling models to construct explicit reasoning chains and execute actions via external resources, such as 3rd Generation Partnership Project (3GPP) APIs exposed by CN NFs. Thus, these resources, which may be referred to as tools, extend the action space of the models and give rise to Agentic AI applied to the CN, in which multiple specialized agents collaboratively perform complex tasks through coordinated reasoning and multi-step planning.

% Limitations and Research Problem Addressed in This Paper

Despite \cite{agentran} and \cite{adv_arch} applying agentic AI in 6G networks, their primary focus is limited to the radio access network (RAN). Moreover, research efforts addressing the xGC remain incipient, and there is a lack of practical demonstrations validating the proposed theoretical solutions. Consequently, a clear gap persists in the literature concerning AI-driven closed-loop mechanisms for real-time self-optimization, self-organization, and self-adaptation in xGC architectures.

Therefore, to meet the complex demands of xGC and given the heterogeneity and dynamism of users, this work proposes Agent\textit{xG}Core, an Agentic AI-native solution for 6G and xGC, positioned within the intelligent layer. This solution aligns with and extends the 3GPP architecture by introducing a framework that enables interactions across xGC domain. The proposed solution is based on a multi-specialized agent architecture, comprising a network planner agent, capable of visualizing the network state and developing plans to satisfy the intents, and a network executor agent, responsible for validating these plans and executing the necessary actions to fulfill them. The contributions of this paper are summarized as follows:

\begin{enumerate}
    \item An overview of the AI4NET paradigm and the state of the art, highlighting the evolution of artificial intelligence models applied to networks and the role of Agentic AI.
    
    \item A proposed xGC architecture that incorporates Agentic AI for the creation of an intelligent layer capable of promoting autonomy and flexibility in the network management, orchestration, and operation. This architecture enables the adoption of the IBN paradigm, which, together with an intent manager, facilitates the formation of AI-driven closed loops.
    
    \item The proposal is assessed using various real data flows, such as Video on Demand (VoD), Live Streaming (LS), and Cloud Gaming (CG). Additionally, a performance evaluation and comparison between different LLM models, such as Gemini and GPT, used as the basis of the proposed architecture, are carried out. 
\end{enumerate}

\section{A Brief Overview of AI/ML in 3GPP}
\label{sec:background}

\begin{figure*}[ht!]
    \centering
    \includegraphics[width=0.85\linewidth]{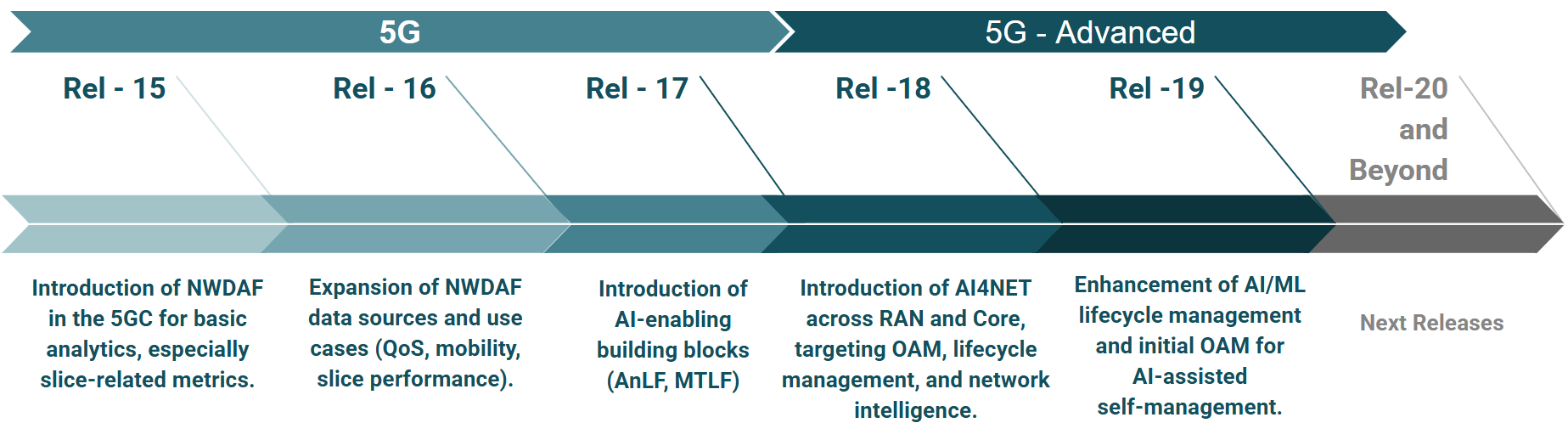}
    \caption{AI/ML evolution over the 3GPP releases for the core network.}
    \label{fig:evolution_3gpp}
\end{figure*}

Figure \ref{fig:evolution_3gpp} illustrates the 3GPP evolution across releases toward an AI-native architecture. Release 15 (Rel-15) introduced the Network Data Analytics Function (NWDAF) in the 5GC to support basic analytics, including network slice monitoring and usage statistics. In Rel-16, NWDAF was extended to incorporate additional data sources and use cases, including enhanced mobility, QoS, and slice performance. At this stage, NWDAF was primarily designed to provide statistical analytics and reporting capabilities rather than advanced intelligence.

In subsequent releases, Rel-17 introduced more advanced AI/ML enablers through the Analytics Logical Function (AnLF) and the Model Training Logical Function (MTLF), allowing standardized support for model training and inference. Nevertheless, despite these enhancements, the NWDAF remained limited from a network orchestration and management perspective. With the introduction of Rel-18 and the advent of 5G Advanced, the 3GPP significantly expanded the role of AI/ML in its specifications, targeting network operation, administration, and maintenance (OAM), and taking the first steps toward the concept of AI4NET in both the RAN and the 5GC, as a foundational step toward AI-native architectures. 

However, Rel-18 primarily introduced these concepts, which were consolidated in Rel-19, mainly through the standardization of AI/ML lifecycle management and AI-assisted self-management capabilities for OAM. Notably, the main advancements remain largely RAN-related. Despite this evolution, the AI/ML system is still primarily focused on decision support rather than on autonomous, closed-loop control. In this context, AI4NET has become a central theme in the evolution toward AI-native architectures, and it is expected that Rel-20 and beyond will further expand these aspects toward more autonomous and intelligent networks.

\section{Evolution of Modern AI Applications in Mobile Networks}

\begin{figure*}
    \centering
    \includegraphics[width=1\linewidth]{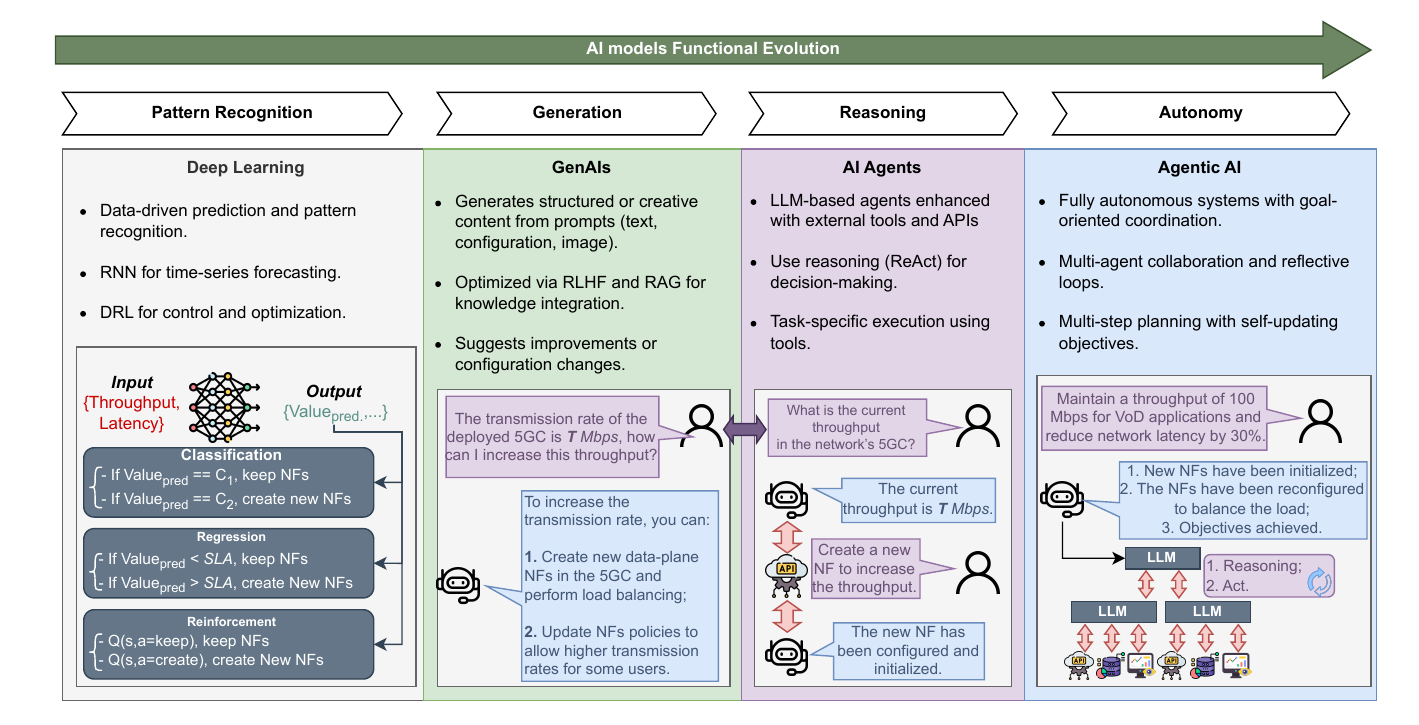}
    \caption{Functional evolution of AI/ML in mobile networks.}
    \label{fig:evolution}
\end{figure*}

Figure \ref{fig:evolution} presents the functional evolution of AI/ML models applied to mobile networks. Early approaches are mainly based on deep learning (DL), with simplified flows that employ numerical inputs and outputs. Classification models, for instance, infer network congestion from metrics such as throughput and latency, indicating the need to scale NFs. Regression models, such as recurrent neural networks (RNN), enable time-series prediction to anticipate SLA violations and support proactive network reconfiguration. In addition, reinforcement learning–based approaches, including single-agent, multi-agent, and federated learning, have been widely adopted for network management and orchestration.

With the advancement of LLMs, a class of transformer-based models driven by generative artificial intelligence, it became possible to use high-level natural language prompts that can understand and generate outputs in text, images, or audio. These models have been applied to mobile networks. 

However, LLMs may exhibit undesired behaviors, such as hallucinations, motivating the adoption of alignment techniques like Reinforcement Learning from Human Feedback (RLHF). In addition, the need to incorporate domain-specific and external knowledge has led to the adoption of Retrieval-Augmented Generation (RAG), enabling LLMs to query external databases and documents before generating responses. As LLM-based systems increasingly require access to external information and participation in decision-making processes, their role extends beyond content generation to action execution and environment observation, giving rise to AI agents and Agentic AI.

An AI agent is an autonomous, reactive entity designed to achieve a specific goal, capable of reasoning from environmental context and taking actions to reach that goal. They can be seen as an evolution of GenAIs, incorporating ReAct capabilities, that is, the ability to construct reasoning chains and execute actions employing tools. These tools correspond to a set of resources, such as APIs, that enable agents to interact with external environments and perform tasks. Tool discovery can occur through direct coupling with the LLM, however, as the system scales, this approach tends to face scalability limitations. In this context, architectures based on Model Context Protocol (MCP) servers \cite{mcp} emerge, acting as brokers that allow different LLMs to operate as clients, thereby expanding the discovery and utilization of available tools.

Despite this, AI agents face limitations in dealing with multiple objectives or complex tasks. Meanwhile, Agentic AI provides an architecture composed of multiple specialized agents that collaborate, are goal-oriented, and capable of multi-step planning to autonomously handle complex activities. This approach enables intent injection, goal-decomposition, reasoning, and coordinated action execution, promoting autonomy, flexibility, self-organization, fault detection and recovery, and continuous intent-based optimization. Moreover, this architecture is not merely reactive, but rather a complex system capable of sharing memory states and collaboratively aligning actions.

\section{Related Work}

This section presents representative related work summarized in Figure \ref{fig:evolution}, focusing on AI approaches applied to xG networks, particularly the CN. Table~\ref{tab:related_work} categorizes those works by their interaction scope. In particular, control and data plane (CP/DP) operation involves direct runtime actions on NFs via 3GPP interfaces, while CN Coordination refers to intent-driven management across CP/UP and MANO.

The approach proposed in \cite{6g_Enablers} integrates intelligent agents into each NF of a 6G core to enable autonomous and distributed decision-making. Although effective for complex decision-support tasks, these models are limited in reasoning and contextual understanding and often introduce overhead due to task-specific agents. The authors in \cite{ibn_5g_deploy} present a LLM-based solution that translates user intents into automated network configurations, and \cite{ibn_service} proposes an intent-driven, autonomous multi-agent system for infrastructure-as-code (IaC) and service orchestration.

Meanwhile, in \cite{ai_agent_ngc}, an autonomous cognitive architecture based on AI agents for the 6G core was proposed, in which agents perceive and act proactively using language models and supporting tools. \cite{ibm_llm} and \cite{oss_gpt} explore architectures based on autonomous agents for managing the intent lifecycle and integrating with network APIs. Regarding mobile network cores, \cite{agentic_core} addresses the growing complexity of the 6GC by proposing an agentic, mission-oriented architecture leveraging GenAI to resolve intents into autonomous workflows. However, the solution remains conceptual. In \cite{agile} is presented a distributed, multi-agent framework operating above traditional MANO systems. Nevertheless, these solution does not demonstrate direct interaction with core NFs, nor closed-loop coordination or management across the xGC control and data planes.

Therefore, the state of the art still lacks solutions capable of realizing complex QoS-driven intents through coordinated intent decomposition and decision-making across the xGC domain, as existing approaches remain primarily confined to management and orchestration layers. To address this gap, this paper proposes an intent-driven coordination framework that enables continuous optimization and closed-loop autonomy by coordinating MANO-level actions with control and data plane operations via standardized 3GPP interfaces, based on operator-defined intents. 

\begin{table*}[ht!]
\centering

\caption{Summary of related works and use of AI Applications}
\label{tab:related_work}
\begin{tabularx}{\textwidth}{c X c c c c c c}
\toprule
\multirow{2}{*}{\textbf{Paper}} &
\multirow{2}{*}{\textbf{Aim}} &
\multirow{2}{*}{\textbf{NFs Management/}} &
\multirow{2}{*}{\textbf{CP and DP}} &
\multirow{2}{*}{\textbf{Coordination}} &
\multirow{2}{*}{\textbf{Closed}} &
\multicolumn{2}{c}{\textbf{AI/ML Used}} \\
\cmidrule(lr){7-8}
 &  & \textbf{Orchestration} & \textbf{Operation} & \textbf{within CN} & \textbf{Loop}  & \textbf{GenAI} & \textbf{Agentic AI} \\
\midrule

\cite{6g_Enablers} &
Intelligent agents integrated into each NF. &
\xmark &
\xmark &
\xmark &
\xmark &
\xmark &
\xmark \\

\midrule
\multicolumn{8}{l}{\textbf{IBN Solutions}} \\
\midrule

\cite{ibn_5g_deploy} &
Configure and deploy 5G networks. &
\checkmark &
\xmark &
\xmark &
\xmark &
\checkmark &
\xmark \\

\cite{ibm_llm} &
Configure and manage network services. &
\checkmark &
\xmark &
\xmark &
\xmark &
\checkmark &
\xmark \\

\cite{ibn_service} &
Multi-agent system executing actions via IaC. &
\checkmark &
\xmark &
\xmark &
\xmark &
\checkmark &
\checkmark \\

\cite{oss_gpt} &
Simplify user interaction and enable autonomous system adaptation for OSS. &
\checkmark &
\xmark &
\xmark &
\xmark &
\checkmark &
\checkmark \\

\cite{agentic_core} &
Reduce 6G complexity through AI-driven, mission-oriented network capabilities. &
\checkmark &
\xmark &
\xmark &
\checkmark &
\checkmark &
\checkmark \\

\cite{agile} &
End-to-end service and network orchestration through coordinated multi-agent systems. &
\checkmark &
\xmark &
\xmark &
\checkmark &
\checkmark &
\checkmark \\

\midrule
\textbf{This work} &
Agentic AI-Native for continuous optimization and autonomous coordination within \textit{xGC} domain. &
\checkmark &
\checkmark &
\checkmark &
\checkmark &
\checkmark &
\checkmark \\

\bottomrule
\end{tabularx}
\end{table*}

\section{Agent\textit{xG}Core: AI-Native solution for xGC}
\label{sec:proposal}

This section describes the architecture of the Agent\textit{xG}Core, which enables autonomous mobile core control through Agentic AI-driven control loops that enable networks to interpret and execute high-level operator intents. The following sections describe the architecture and design principles. 

\begin{figure*}[ht!]
    \centering
    \includegraphics[width=1\linewidth]{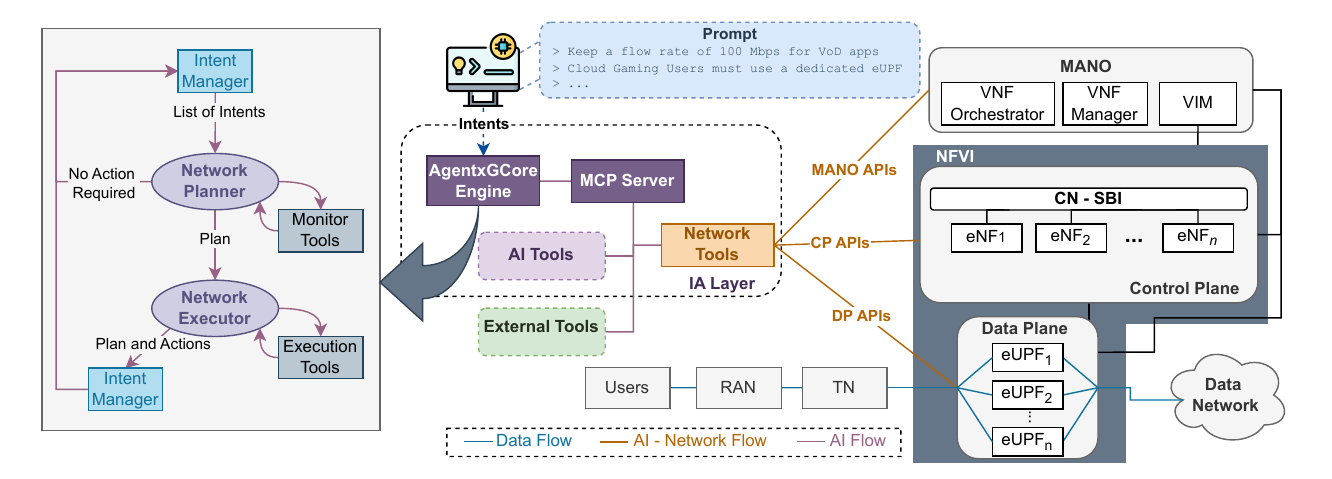}
    \caption{Overview of the architecture of the proposed solution.}
    \label{fig:proposed_solution}
\end{figure*}

\subsection{Architecture Overview}

The Agent\textit{xG}Core is built upon an intelligent layer that extends the 3GPP core architecture, enabling autonomous, AI-driven, closed-loop coordination within the \textit{xGC} domain by jointly operating management functions and CP/DP at runtime. As illustrated in Figure \ref{fig:proposed_solution}, the Agent\textit{xG}Core receives high-level operator intents, processes them through LLMs, and implements the resulting decisions via execution tools, using the APIs defined by 3GPP and MANO. To this end, the MCP Server enables the Agent\textit{xG}Core to interact with the 3GPP architecture by exposing mechanisms for discovering and integrating network-available tools, allowing the required APIs to be identified and invoked. \textcolor{black}{Unlike static rule-based orchestration over these interfaces, which requires hand-coding when, which, and in what order to invoke them and does not generalize to unforeseen conditions, the proposed framework reasons over operator intents to determine the appropriate API sequences.} \textcolor{black}{Finally, agent coordination follows the Agent-to-Agent (A2A) paradigm, with the workflow structured as a graph in which the Planner and Executor exchange structured plans and execution feedback through well-defined message schemas.} The following subsection describes the Agent\textit{xG}Core components.

\subsection{Intent Manager} The intent manager stores and manages the intents. Intents can be classified as one-shot, which require a single action (e.g., creating or deactivating network function instances), and continuous control, which require ongoing management (e.g., ensuring SLAs in a network slice). The Intent Manager organizes these intents and creates a loop of calls to the Network Planner with the actions to be executed at each moment.

\subsection{Network Planner Agent} The agent receives intents and identifies those that require action. It queries the Monitor Tools to obtain the current network state and, based on this information, the LLM evaluates whether the intents are already being met. If not, the LLM develops an action plan, defining steps and methodology, which are then passed on to the Network Executor.

\subsubsection{\textcolor{black}{Monitor Tools}}\textcolor{black}{build the network context consumed by the agent. They rely primarily on \textit{Network Tools}, observability interfaces that combine 3GPP APIs, exposing QoS metrics such as NWDAF data on traffic volume in Protocol Data Unit (PDU) sessions or the number of active sessions in a UPF, and MANO APIs based on standards such as Prometheus and Kubernetes, which expose VNF state and computational resource usage (CPU, RAM). Beyond these, Monitor Tools also incorporate \textit{AI Tools}, additional models and agents that support the Agentic AI Engine, such as DL models for traffic prediction, used here to anticipate whether the intent will remain satisfied. Finally, they include \textit{External Tools}, sources of contextual information that enrich the AI Tools and the LLM, such as regional event calendars or edge computing services that report end-to-end metrics (e.g., latency in cloud gaming or throughput in VoD).}

\subsection{Network Executor Agent} The agent assesses the feasibility of the action plan and, through the MCP Server, identifies the most suitable tools to execute it. It then triggers these tools, which implement the actions in the network. \textcolor{black}{The Executor maintains a contextual memory of executed actions and observed network states, structured as a bounded sliding window to control context size as the deployment evolves.}

\subsubsection{\textcolor{black}{Execution Tools}}\textcolor{black}{perform actions on the xGCore components, relying primarily on \textit{Network Tools} that combine MANO APIs, enabling management, orchestration, and configuration of VNFs, including the creation of new NFs, and 3GPP APIs, which enable resource allocation, user policy updates in the Policy Control Function (PCF) and session modifications in the Session Management Function (SMF). This design allows the agent to operate on slow-timescale orchestrations, building a reasoning chain of sequential actions and then offloading the resulting policies to the CP/DP so they execute independently, avoiding latency aggregation in CP/DP operations. Beyond these, Execution Tools also incorporate \textit{AI Tools} to optimize execution decisions, for example through DL-based load balancing across eUPFs.}
\section{Case Study}
\label{sec:eval}

To demonstrate the feasibility and functionality of the proposed solution, and to quantify its efficiency and validate whether it provides the necessary support for emerging applications, an evaluation was conducted across heterogeneous data flows under different user loads and LLM models.

\subsection{Scenario}

A network composed of diverse data traffic flows and subject to distinct user loads faces the challenge of allocating users coexisting on the same network substrate, especially in the data plane. Defining the optimal number of UPFs while ensuring their full utilization without causing overload or underutilization is a relevant challenge. Furthermore, methodologies based on PDU session transfer can increase latency and lead to data loss. For this reason, it is important to create a planned environment that can self-organize before the arrival of the traffic load. \textcolor{black}{These intents express QoS-driven goals over the DP, with the specific threshold values representing operator-defined parameters that, in production deployments, would be derived from SLA contracts}. To this end, three intents were established:

\begin{enumerate}
    \item \textit{Keep all the user plane network function (UPF) Downlink usage around 25 MB/s}.
    \item \textit{Ensure that no more than 3 UPF instances exist at any time; create new UPFs only as demand increases, and distribute users among the available instances}.
    \item \textit{Stop any UPF instances that are underutilized and currently do not serve any users. But ensure that at least one UPF instance is available}.
\end{enumerate}

\subsection{Experimental Setup}

The Agent\textit{xGC}ore was implemented as an extension of the OpenAirInterface (OAI) 5G Core. For this purpose, a Docker-based containerized environment was used, enabling the creation of an API to activate and deactivate CN function nodes and providing access to orchestration and management functionalities. For control-plane configuration management, 3GPP APIs for the PCF and SMF were exposed, enabling visualization of user data flows, policy updates, and user redistribution across different UPFs, as well as observation of the context and number of PDU sessions in each instance.

To provide the Agent\textit{xGC}ore with a comprehensive view of the network, tools integrated with Prometheus APIs were created, enabling verification of the number of available network function instances and their utilization. To enable proactive self-organization, an AI tool for traffic load prediction was developed based on a Gated Recurrent Unit. \textcolor{black}{The model takes as input a time series of downlink traffic and predicts the load over a 10-second horizon. With this information, the Agent\textit{xGC}ore decides whether to create a new UPF and allocate the still-unconnected users to it or maintain the network in its current state.}

For the radio access network and users, the \texttt{gnbsim} tool was used with 21, 34, 55, and 89 users, assuming an arrival rate of 5 per user per second. The data flows from the dataset \cite{ref12} used includes: Cloud Gaming, which requires low latency and has a high download rate due to video streaming; Live Streaming, which is not latency-critical and has high throughput due to batch video transmission; and Video on Demand, characterized by high traffic peaks but without low-latency requirements. The data were summarized and injected into the network using the \texttt{iperf3} tool.

Finally, in order to assess the impacts of the trade-off between a  number of parameters of a model and its response speed, we used two versions of Gemini, Pro 2.5 and Flash 2.5, and of GPT, 4.1 and 4.1 Mini, as the LLMs of the Agent\textit{xGC}ore.

\subsection{Experimental Results}

The evaluation considered three metrics: (i) execution time, which corresponds to the sum of the total processing time within the AgenticEngine, that is, the total time spent on LLM calls and the interactions between the LLM and tools; (ii) the average downlink throughput, to verify how close each model came to fulfilling the intent; (iii) and the round-trip time (RTT) resulting from the decision-making process of creating UPFs and distributing users among the instances.

Figure~\ref{fig:evolution}a presents the results for the average execution time. It can be observed that the GPT family models achieved the lowest times, indicating underutilization of tools and suggesting difficulty in understanding and planning the actions required to fulfill the intent. This behavior is reflected in Figure~\ref{fig:evolution}b, where these models deviate from the target value indicated by the dashed line.

\begin{figure}[ht!]
    \centering
    \includegraphics[width=1\linewidth]{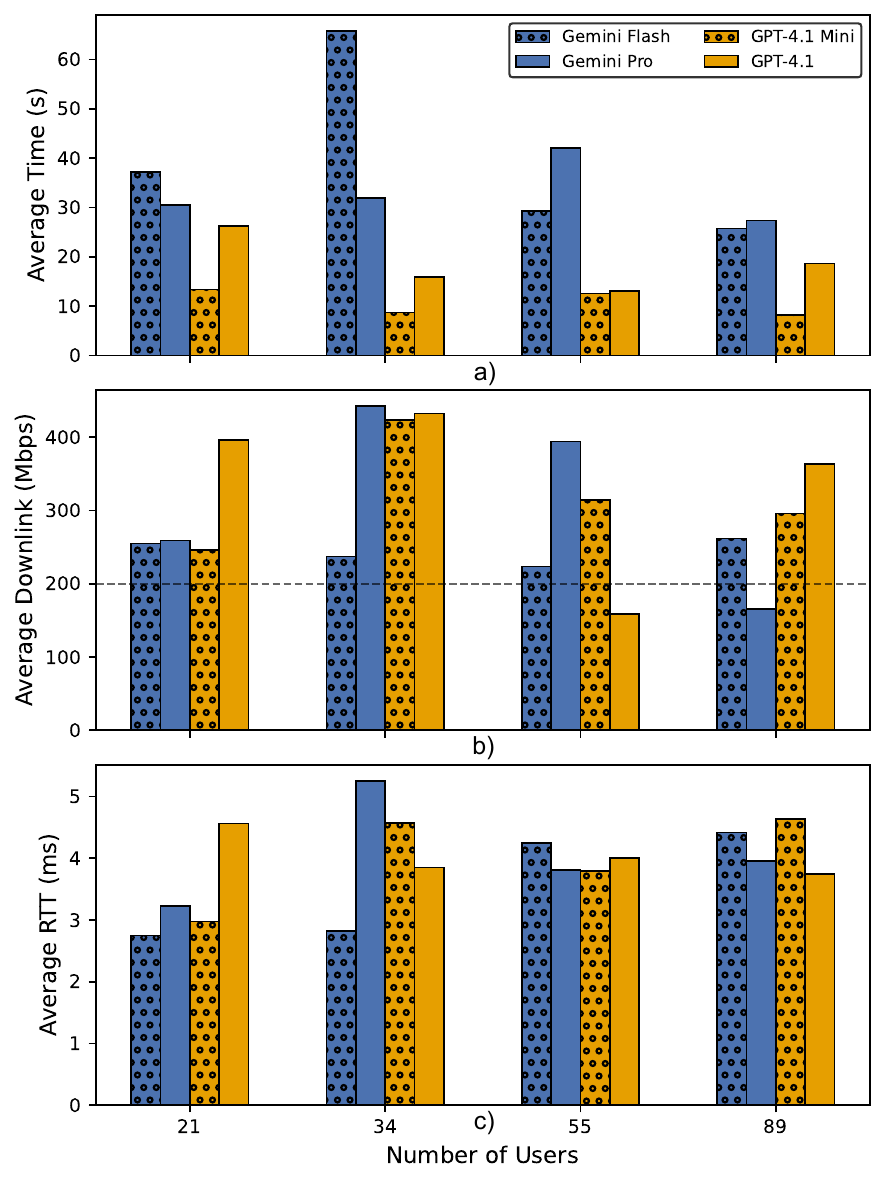}
    \caption{Results of the use of the Agent\textit{xGC}ore on the network in terms of a) Average Execution Time, b) Average Downlink, and c) Average Round Trip Time between the user and the iPerf server.}
    \label{fig:results}
\end{figure}

It is particularly noticeable that models with more parameters, such as GPT~4.1 and Gemini~Pro~2.5, struggle with simple, continuous tasks, often overthinking. This is also evident in Figure~\ref{fig:evolution}c, especially in scenarios with fewer users, indicating ineffective distribution and the creation of instances that do not reduce RTT.

On the other hand, results from Gemini~Flash indicate that simpler models can be more effective for continuous automation activities. Due to the fewer parameters, the model maintains a more stable interpretation of the surrounding context and efficiently self-learns from previous states, reinforcing its suitability for implementing the proposed solution. Furthermore, in scenarios with fewer users, Gemini~Flash shows longer execution time, demonstrating that, in addition to being accurate in planning, it utilizes available resources to meet the intents, positively impacting both throughput, which remained close to the required value, and RTT control.

\section{Open Research Issues}

\subsection{Security}  
During the experiments, it was possible to observe differences in the decision-making process between models, highlighting the importance of context for LLMs. When generic models are used, failures in planning formulation and hallucinations may occur, making the solution ineffective or requiring more time to converge to an appropriate action. Thus, granting the Agentic AI full control of the network may lead to continuous activations and deactivations of NFs and PDU session interruptions. Furthermore, the use of remote models requires sending contextual network information to external servers, which may expose sensitive data such as internal policies, performance metrics, and user information, increasing susceptibility to attacks. Therefore, it is essential to adopt prompt engineering techniques, limit tool functionality, implement human-in-the-loop mechanisms, perform fine-tuning, and use local models with anonymization and encryption to ensure data security and privacy.

\subsection{Local versus remote LLM servers}  
\textcolor{black}{The execution times reported in Section VI-C, which include LLM inference and tool interaction, illustrate the trade-offs of relying on external LLMs. Increased latency was observed during querying and response processing.} Additionally, there are issues related to LLM server availability, as interruptions or request limits can affect the network’s ability to respond in real time. Therefore, it is essential to use local LLM servers, such as Llama, which support multiple LLM models. However, many of these models require specialized hardware to operate properly; otherwise, even simple tasks may exhibit high processing latency when executed locally. Moreover, there is the computational cost of keeping the model running continuously and periodically updating it through fine-tuning. Hence, hybrid strategies that combine local and remote models become necessary.

\subsection{Messages Overhead}  
An increase in data traffic in Internet backbone is expected due to the flow of information exchanged with LLMs. Additionally, continuous monitoring incurs API call overhead, as it is required to constantly check the network state. This increase in traffic may affect system scalability and require prioritization or message compression, especially in environments with multiple agents or distributed functions.

\section{Conclusion}
\label{sec:conc}
This paper proposes Agent\textit{xG}Core, an Agentic AI-native solution composed of multiple specialized agents that enable intent-driven coordination within the \textit{xGC} domain, operating in a closed loop and jointly supporting management and orchestration functions alongside direct control and data plane operations at runtime, based on operator-defined intents. To enable this approach, the 3GPP architecture is extended to reuse APIs already standardized in 5G, which are registered in an MCP server and made available to agents in the intelligent layer. The proposed solution is validated using an open-source core network and real traffic datasets. The evaluation considers different LLMs, including Gemini Flash, Gemini Mini, GPT-4.1, and GPT-4.1 Mini, focusing on network load-balancing scenarios. The results demonstrate the effectiveness of the proposed solution, indicating that models with fewer parameters are better able to satisfy the defined objectives, highlighting the potential of the proposed approach for autonomous management and orchestration in the xGC. \textcolor{black}{As future work, beyond the open research issues already discussed, cross-domain coordination between the Core and RAN remains a key direction toward fully autonomous end-to-end network operation and exploring higher-level goal-oriented intents.}

\section*{Acknowledgment}
This work was supported by the National Council for Scientific and Technological Development (CNPq) - Research Productivity Fellowship (Grant No. 313083/2023-1) and Pernambuco Research Foundation (FACEPE) (Grant No. IBPG-0130-1.03/23).

%%%%%%%%%%%%%%%%%Referencias%%%%%%%%%%%

\vspace{20pt} 
\footnotesize
\textbf{Maria Katarine Santana Barbosa} is a Ph.D. student in Computer Science at the Center of Informatics of the Federal University of Pernambuco (CIn/UFPE). Her current research interests are Optimization, Network Functions Virtualization, Software-defined Radio, 5G and 6G networks.

\textbf{Kelvin Lopes Dias} is a full professor at the Center of Informatics (CIn) of the UFPE. His current research interests are Software-defined Networking, Network Functions Virtualization, Edge Computing, 6G networks, and Advanced \& Intelligent Network Architectures.

\end{document}